\documentclass[11pt]{article}

\usepackage[preprint]{acl}

\usepackage{times}
\usepackage{latexsym}

\usepackage[T1]{fontenc}

\usepackage[utf8]{inputenc}

\usepackage{microtype}

\usepackage{inconsolata}

\usepackage{graphicx}
\usepackage{subcaption}
\usepackage{booktabs, multirow}
\usepackage{amsmath, amssymb}
\usepackage{xcolor}
\newcommand{\val}[2]{#1$_{\,\textcolor{gray}{#2}}$}                  
\newcommand{\valb}[2]{\textbf{#1}$_{\,\textcolor{gray}{#2}}$}        
\newcommand{\valu}[2]{\underline{#1}$_{\,\textcolor{gray}{#2}}$}     

\title{Neighborhood-Aware Dual Biomedical Entity Linking}

\author{
 \textbf{Yicheng Tao\textsuperscript{1}},
 \textbf{Jie Liu\textsuperscript{1,2}}
\\
 \textsuperscript{1}Department of Electrical Engineering and Computer Science, University of Michigan
\\
 \textsuperscript{2}Gilbert S. Omenn Department of Computational Medicine and Bioinformatics, University of Michigan
}

\begin{document}
\maketitle
\begin{abstract}
Biomedical entity linking grounds mentions in clinical and scientific text to entities in a curated knowledge base (KB) with ontological structure, which supports downstream applications such as literature-scale information extraction and patient-record normalization. The task has several challenges at once: the KB contains large numbers of entities, mentions are often ambiguous, and gold labels follow annotation conventions specific to each corpus. To address these challenges, we propose PILOT, a three-stage framework made up of neighborhood-aware retrieval, dual reranking, and score fusion. The retriever injects ontological structure from both the query and KB side, by reformulating mentions and pooling entity embeddings. The retrieved pool is then scored from two complementary views, one over surface forms and one over context, and fused together. PILOT achieves the state of the art on average across five widely-used benchmarks and remains efficient at inference. 
\end{abstract}

\section{Introduction}

Biomedical entity linking (BioEL) grounds mentions in clinical and scientific text to entities in a curated knowledge base (KB), such as UMLS \citep{bodenreider2004unified}. These KBs are typically organized as ontologies, in which each entity carries a set of synonyms and a position in an ``is-a'' hierarchy. BioEL underpins downstream applications from literature-scale information extraction \cite{huang2024building} to patient-record normalization \cite{tao2026autopcr}. Its difficulty comes from three challenges.
The first is scale. A target KB may contain millions of entities, so exhaustive scoring is infeasible and the search space has to be narrowed before any expensive per-candidate computation. 
The second is ambiguity. Biomedical mentions are riddled with abbreviations, informal paraphrases, and terms whose surface forms are nearly identical but whose referents are not, so lexical similarity alone cannot separate the gold entity from its neighbors. The surrounding sentence is frequently the only signal that disambiguates. 
The third is annotation. When no entity carries exactly the mention's meaning, annotators fall back on the closest available one, and the resulting conventions are specific to each corpus and KB. A system that never sees in-domain supervision cannot reproduce them.

Existing methods approach these challenges from different paradigms, each with its own limitations. Discriminative encoders \citep{sung2020biomedical, liu2021self, lai2021bert} scale to the full KB through dense retrieval, but score a mention against an entity name in isolation and are therefore weakest exactly where context matters. Generative linkers \citep{yuan2022generative, kim2025learning} decode entity names directly under a constrained vocabulary and absorb annotation conventions through fine-tuning, but the ranking signal is a sequence likelihood rather than a calibrated measure of relevance. LLM-based methods \citep{lin2025guiding} compare retrieved candidates in one prompt, but this caps the candidate set and exposes the ranking to context-window limits, positional attention decay \citep{liu2024lost}, and sensitivity to candidate order \citep{sun2023chatgpt}.
To overcome these limitations, we propose PILOT, a three-stage framework made up of neighborhood-aware retrieval, dual reranking, and score fusion. The retriever is made neighborhood-aware from two directions at once. On the query side, an instruction-tuned LLM rewrites each mention into a standardized name and a hypernym, which are mixed with the mention embedding. On the KB side, each entity embedding is pooled over its parents and children by a one-hop Rocchio \citep{rocchio1971relevance} update. The resulting pool is then scored from two complementary views: a precision-configured second pass of the same encoder focusing on surface forms, and a pointwise generative cross-encoder that reads the mention in its context. Scoring candidates one at a time keeps the input length constant in the size of the pool, which lets PILOT rerank a much deeper pool than prompt-based methods afford. The two scores are finally fused with a single weight, so that lexical and contextual evidence are combined rather than chosen between.

Our contributions are threefold. First, we propose a neighborhood-aware retriever that injects ontological structure symmetrically, through generative query reformulation on the query side and ontology-based embedding pooling on the KB side. Second, we propose a dual reranker that scores candidates from a surface-form and a contextual view and fuses the two calibrated scores with a single weight, trained with hard negatives mined from both views. Third, PILOT achieves the best result on four of five BioEL benchmarks and the highest average overall, while running an order of magnitude faster than the strongest LLM-based baseline.


\section{Related Work}

\paragraph{Biomedical entity linking.}
BioEL methods fall into three categories: discriminative, generative, and LLM-based. Discriminative approaches embed mentions and entity names in a shared space and link by nearest-neighbor search. BioSyn \citep{sung2020biomedical} treats the synonyms of an entity as equivalent targets and marginalizes over them during training, with candidates refreshed iteratively as the encoder improves. SapBERT \citep{liu2021self} pretrains a biomedical encoder by self-alignment over UMLS synonym pairs and has since become a standard retrieval backbone. ResCNN \citep{lai2021bert} shows that a lightweight residual convolutional encoder matches far larger transformers on this task. KRISSBERT \citep{zhang2022knowledge} injects context sensitivity by generating self-supervised mention examples from a KB rather than relying on annotated corpora. Prompt-BioEL \citep{xu2023improving} departs from independent scoring by placing the retrieved candidates in a shared prompt so that they interact, at the cost of a small candidate set. Generative approaches instead decode the entity name directly. Following the general-domain precedent of autoregressive entity retrieval \citep{decao2021autoregressive}, GenBioEL \citep{yuan2022generative} decodes under a prefix-tree constraint derived from the KB, combining KB-guided pretraining with synonyms-aware fine-tuning on backbones such as BART \citep{lewis2020bart} and BioBART \citep{yuan2022biobart}. ANGEL \citep{kim2025learning} observes that such models are trained only on positives, and adds a negative-aware stage in which the model's own incorrect top-$k$ generations, together with TF-IDF-similar names, are used for preference optimization. Most recently, general-purpose LLMs have been applied directly: GPT-4 \citep{achiam2023gpt} and DeepSeek-R1 \citep{guo2025deepseek} are used zero-shot, while LLM4BioEL \citep{lin2025guiding} improves them by including retrieved candidates and few-shot examples in the prompt, with restrictive decoding over KB-valid names and contrastive decoding against a context-free prior. PILOT belongs to this last category, but uses the LLM to rerank a deep candidate pool rather than to generate the answer, and replaces the single reranking pass with two views that are scored separately and fused.

\paragraph{Query reformulation.}
Rewriting the query before retrieval dates back to Rocchio's relevance feedback \cite{rocchio1971relevance}, and has since been applied wherever the user's input is an imperfect proxy for the information need. In conversational search, rewriting resolves the coreference and ellipsis left by the dialogue history \cite{wu2022conqrr, mao2023large}. In retrieval-augmented generation, it aligns the question with what the retriever expects \cite{ma2023query, mao2024rafe}. In web and e-commerce search, it bridges long-tail queries to the product vocabulary \cite{peng2024large}.
Within retrieval itself, modern reformulations are usually generative. Query2doc \cite{wang2023query2doc} prompts an LLM for a pseudo-document and appends it to the query, HyDE \cite{gao2023precise} embeds a generated document and searches with it in place of the query, and CSQE \cite{lei2024corpus} expands the query with key sentences taken from an initial retrieval. These reformulations supply context that a short query lacks, closing the lexical gap to the documents that answer it.

\paragraph{Reranker.}
Reranking a retrieved pool with a cross-encoder was established by monoBERT \cite{nogueira2019multi}, which jointly encodes a query--document pair, and monoT5 \cite{nogueira2020document}, which recasts the same computation as generation and reads relevance off the logit of a single decision token. RankT5 \cite{zhuang2023rankt5} later fine-tunes the same architecture directly with ranking losses. 
Prompting LLMs to rerank has since opened a broader design space, usually organized into pointwise, pairwise, listwise, and setwise paradigms \cite{zhuang2024setwise}, which differ in how many candidates enter a single prompt and therefore in the trade-offs they make.
Pointwise rerankers \cite{zhuang2024beyond, zhang2025qwen3} judge one candidate at a time, so the prompt length is independent of the pool size and every candidate receives a score that does not depend on the rest of the pool. The price is that the model never sees two candidates side by side, making near-identical candidates hard to rank, which is usually mitigated by training with hard negatives.
Pairwise rerankers \cite{qin2024large, luo2024prp} compare two candidates at a time, which gives a fine-grained relative signal, but need a super-linear number of comparisons to sort a pool, and their comparisons may violate transitivity.
Listwise rerankers \cite{sun2023chatgpt, pradeep2023rankzephyr} see many candidates at once and output a permutation in a single pass, but they can only fit a limited number of candidates in one prompt and must slide a window over longer lists. Their output is also sensitive to the order in which candidates appear, an effect compounded by the tendency of long-context models to attend less to material in the middle of their input \cite{liu2024lost}.
Setwise rerankers \cite{zhuang2024setwise} sit between the two, selecting the most relevant candidate from a small set at each step, which cuts the number of inference calls needed to sort a pool while retaining direct comparison.

\begin{figure*}[t]
  \centering
  \includegraphics[width=\textwidth]{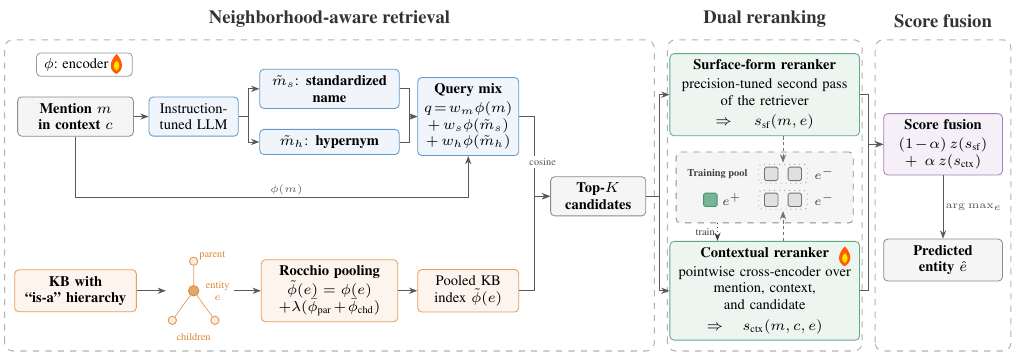}
  \caption{Overview of PILOT.}
  \label{fig:overview}
\end{figure*}

\section{Method}

Biomedical entity linking (BioEL) maps a mention $m$ in context $c$ to an entity in a knowledge base (KB) $\mathcal{K}=\{e_1,\dots,e_{|\mathcal{K}|}\}$, where each entity $e$ carries a set of synonyms and a position in an ontology (``is-a'' hierarchy). The task is challenging on three fronts simultaneously: the KB is large (up to $3.1\text{M}$ entities / $6.1\text{M}$ synonyms, UMLS); lexical matching is insufficient since isolated mentions are often ambiguous and need context to be resolved; and a gold entity with exactly the same meaning as the mention may not exist, in which case the ``closest'' one is chosen, introducing dataset-specific annotation conventions.
PILOT addresses these challenges with three stages (Figure~\ref{fig:overview}): a \textbf{neighborhood-aware retriever} that scales to the entire KB and gathers ontologically adjacent entities into a high-recall candidate set; a \textbf{dual reranker} that scores those candidates from two complementary views---a \textit{surface-form reranker} that sharpens lexical precision and a \textit{contextual reranker} that disambiguates mentions using their context; and a \textbf{score fusion} step that fuses the two into the final ranking. PILOT's design is decoupled: retrieval maximizes recall, reranking maximizes precision, and both stages are trained to capture the dataset's annotation conventions.

\subsection{Neighborhood-aware retrieval}
The retriever encodes the query mention and every KB synonym with a fine-tuned SapBERT encoder\footnote{SapBERT is fine-tuned per dataset using its original training recipe based on contrastive learning, and we use its fine-tuned version throughout the paper.} and returns the top-$K$ unique entities by cosine similarity. We include two design elements to make it ``neighborhood-aware'' so as to increase recall.

\paragraph{Generative query reformulation.}
The query mention may not be in an accurate form for retrieval, since parts of its semantics may be distributed in the context or it may be too specific for a KB to match. Thus, we reformulate it with an instruction-tuned LLM that takes the mention and its context, and emits, per mention, the most appropriate standard scientific/medical name (``standardized name'') and the broader parent category or general concept it is a kind of (``hypernym''). The prompts are given in Appendix~\ref{app:query_prompts}. The query embedding is then a convex combination
\begin{equation}
\begin{aligned}
    q &= w_{m}\,\phi(m) + w_{s}\,\phi(\tilde m_{s}) + w_{h}\,\phi(\tilde m_{h}), \\
    &\quad \text{with}\quad w_m+w_s+w_h=1,
\end{aligned}
\end{equation}
where $\tilde m_{s}$ (standardized name) and $\tilde m_{h}$ (hypernym) are the LLM reformulations. This injects the ``ontological neighborhood'' on the query side: the standardized name pushes the mention toward its child entities to capture the specificity expressed in the context and the hypernym pulls the mention toward its broader parent entities when the mention itself is too specific.

\paragraph{Ontology-pooled KB embeddings.}
On the KB side we apply the complementary enrichment: each entity embedding is pooled over its ``is-a'' neighborhood by a one-hop Rocchio \cite{rocchio1971relevance} update over parent and child entities,
\begin{equation}
    \tilde\phi(e) = \phi(e) + \lambda\left(\bar\phi_{\text{par}}(e) + \bar\phi_{\text{chd}}(e)\right),
\end{equation}
where $\bar\phi_{\text{par}},\bar\phi_{\text{chd}}$ average the embeddings of the parents and children. Pooling makes an entity retrievable through its neighbors, recovering mentions whose surface form matches a sibling or parent more closely than the gold string.

The two enrichments share the same principle---injecting the ontological neighborhood, one on the query side and one on
the KB side---but are mechanically distinct (generative reformulation vs. embedding pooling). Together, they improve the recall of the retriever, with the query-mix weights $(w_m,w_s,w_h)$ and pooling weight $\lambda$ tuned on the development set to maximize Recall@$K$ (R@$K$).

\subsection{Dual reranking}
The dual reranker scores the $K$ retrieved candidates from two complementary views: a surface-form view that sharpens string-level discrimination, and a contextual view that exploits the mention's context and the candidate's ontological neighborhood.

\paragraph{Surface-form reranker.} The surface-form reranker is a second pass of the neighborhood-aware retriever, but re-scores only the $K$ candidates and is configured for precision rather than recall, via a sharper query mix and pooling tuned on development by R@1. It remains a string-similarity model, hence \emph{surface-form}, but the precision configuration separates the true entity from the near-lexical neighbors that the recall-tuned retriever deliberately blurs together. Because it only re-ranks the retriever's $K$ candidates, this sharpness costs nothing in recall. Its score for entity $e$ is $s_{\text{sf}}(m,e)=\cos(q^{\text{sf}},\tilde\phi^{\text{sf}}(e))$.

\paragraph{Contextual reranker.} The contextual reranker is a generative cross-encoder that reads the mention in its context together with the candidate's synonyms and a one-hop textual rendering of its ontological neighborhood (parents, children, and siblings; see prompts in Appendix~\ref{app:reranker_prompt}). We hypothesize that this context- and knowledge-aware view resolves cases where the surface string is ambiguous but the sentence or the ontological context is disambiguating.
We adopt a pointwise reranker rather than a pairwise, listwise, or setwise one for two reasons. First, a pointwise reranker assigns each candidate an individual score, which can later be fused with the score produced by the surface-form reranker. Second, it scales gracefully to a large $K$: candidates are scored independently with constant-length inputs, without the super-linear comparison cost \citep{qin2024large} of pairwise rerankers, or the context-window constraints, positional attention decay \citep{liu2024lost}, and candidate-order sensitivity \citep{sun2023chatgpt} of listwise and setwise rerankers.
The reranker emits a relevance score $s_{\text{ctx}}(m,c,e)$ defined as the yes/no logit margin at the answer position. We train it with a pointwise binary cross-entropy loss over yes/no labels, using the hard negatives introduced below.

\paragraph{Hard negatives.} The contextual reranker is trained on the retriever's $K$-candidate pool. For each positive pair $(m,e^{+})$, we mine $n$ hard negatives from this pool via two complementary sources: (i) a \emph{surface-form semi-hard band}---non-gold candidates whose surface-form score falls just below that of the gold entity, i.e., within $[\,s_{\text{sf}}(m,e^{+})-\delta_{\text{lo}},\; s_{\text{sf}}(m,e^{+})-\delta_{\text{up}}\,]$, where the upper margin $\delta_{\text{up}}$ discards likely false negatives (candidates scored at or above the gold) and the lower margin $\delta_{\text{lo}}$ discards trivially easy ones; and (ii) the reranker's own top-ranked non-gold candidates, so that the model learns to suppress the errors it is most prone to. We draw at most $n_\text{sh}$ negatives from the semi-hard band and fill the remainder with the reranker's top non-gold candidates, up to $n$ per positive. The two sources are complementary: (i) supplies confusable lexical neighbors, while (ii) supplies confusable contextual ones.

\subsection{Score fusion}
We combine the complementary reranker scores rather than choosing between them. For each query we $z$-normalize both score vectors over the $K$ candidates and take a convex combination,
\begin{equation}
\begin{aligned}
    s(m,c,e) &= (1-\alpha)\,z(s_{\text{sf}}) + \alpha\,z(s_{\text{ctx}}), \\
    \hat e &= \arg\max_{e}\,s(m,c,e),
\end{aligned}
\end{equation}
with a single fusion weight $\alpha{\in}[0,1]$ selected on development by R@1. The $z$-normalization puts the two heterogeneous scores on a common scale so that $\alpha$ trades them off directly; $\alpha{=}0$ recovers surface-form ranking, and $\alpha{=}1$ contextual ranking.


\section{Experiments}

\subsection{Experimental setup}
We adopt the same evaluation protocol as ANGEL \cite{kim2025learning}, considering five widely-used BioEL benchmarks: NCBI \cite{dougan2014ncbi}, BC5CDR \cite{li2016biocreative}, COMETA \cite{basaldella2020cometa}, AAP \cite[AskAPatient]{limsopatham2016normalising}, and MM-ST21pv \cite{mohan2019medmentions}.
NCBI comprises disease mentions from 793 PubMed abstracts, normalized to MEDIC \cite{davis2012medic}, an ontology that merges disease entities from MeSH \cite{lipscomb2000medical} and OMIM \cite{hamosh2005online}.
BC5CDR extends the coverage to both disease and chemical mentions across 1,500 PubMed abstracts, linked against MeSH.
Shifting from scientific literature to social media, COMETA collects biomedical mentions from 68 health-related subreddits, while AAP draws on 8,662 social media phrases; both are grounded in SNOMED CT \cite{chang2021use}.
Finally, MM-ST21pv spans 21 selected UMLS \cite{bodenreider2004unified} entity types over 4,392 PubMed abstracts, offering the broadest semantic scope.
Detailed dataset statistics are summarized in Table~\ref{tab:dataset_stats}.
We report test Recall@1 (R@1, \%), also known as top-1 accuracy in our setting, as the evaluation metric. Since AAP provides no official test split, we instead perform 10-fold cross-validation and report the average performance.

\subsection{Baselines}
We compare PILOT against state-of-the-art BioEL baselines from three categories: (1) discriminative methods: BioSyn \cite{sung2020biomedical}, SapBERT \cite{liu2021self}, ResCNN \cite{lai2021bert}, KRISSBERT \cite{zhang2022knowledge}, and Prompt-BioEL \cite{xu2023improving}; (2) generative methods: BART \cite{lewis2020bart}, BioBART \cite{yuan2022biobart}, GenBioEL \cite{yuan2022generative}, and ANGEL \cite{kim2025learning}; and (3) LLM-based methods: GPT-4 \cite{achiam2023gpt}, DeepSeek-R1 \cite{guo2025deepseek}, and LLM4BioEL \cite{lin2025guiding}. For each method, we adopt its best-performing variant. The results of the discriminative and generative methods are directly taken from the ANGEL paper, and those of the LLM-based methods are from the LLM4BioEL paper.

\subsection{Implementation details}
We use Qwen3-4B-Instruct \cite{yang2025qwen3} for generative query reformulation and Qwen3-Reranker \cite{zhang2025qwen3} for contextual reranking, with the 0.6B variant for large-scale MM-ST21pv and the 4B variant for the others. We fine-tune the reranker with LoRA \cite{hu2022lora}, targeting all linear modules with rank $16$ and scaling factor $32$, for $E{=}2$ epochs, using a cosine-decayed learning rate of $5{\times}10^{-5}$ with warmup ratio $0.05$, batch size $8$, and maximum sequence length $512$. All LLMs are decoded greedily.
We select the query-mix weights $(w_m,w_s,w_h)$ over a grid of $(w_m,w_s{:}w_h){\in}\{0.3,0.4,\dots,1.0\}\times\{1{:}0,1{:}1,0{:}1\}$ and the pooling weight $\lambda{\in}\{0,0.1,\dots,0.5\}$ on the development set; exact values are shown in Figures~\ref{fig:sensitivity_mix} and~\ref{fig:sensitivity_pool} and Appendix~\ref{sec:details_aap}. We mine $n{=}8$ negatives per positive from the $K{=}50$-candidate pool, and the semi-hard band uses an upper margin $\delta_{\text{up}}{=}0.02$ and a lower margin $\delta_{\text{lo}}{=}0.30$, from which we take up to $n_\text{sh}{=}5$ negatives. We select the fusion weight $\alpha{\in}\{0,0.05,\dots,1.0\}$ on the development set, with exact values in Figure~\ref{fig:sensitivity_alpha}.
We conduct all experiments on a compute node with an Intel Xeon Gold 6242R CPU (8 cores allocated), 128\,GB RAM, and two NVIDIA H100 GPUs (80\,GB VRAM each). Training takes at most 6 hours per dataset (longest on MM-ST21pv).

\begin{table*}[t]
    \centering
    \resizebox{\ifdim\width>\textwidth \textwidth \else \width \fi}{!}{
        \begin{tabular}{ll|ccccc|cc}
\toprule
\multicolumn{2}{c|}{\textbf{Method}} & \textbf{NCBI} & \textbf{BC5CDR} & \textbf{COMETA} & \textbf{AAP} & \textbf{MM-ST21pv} & \textbf{Avg. (first 3)} & \textbf{Avg.} \\
\midrule
\multirow{5}{*}{Discriminative}
 & BioSyn       & 91.1 & 93.3 & 71.3 & 86.5 & --   & 85.2 & --   \\
 & SapBERT      & 92.3 & 88.6 & 75.1 & 89.0 & 50.3 & 85.3 & 79.1 \\
 & ResCNN       & 92.4 & 94.0 & 80.1 & 77.4 & 55.0 & 88.8 & 79.8 \\
 & KRISSBERT    & 91.3 & 72.0 & 80.1 & 83.1 & 72.2 & 81.1 & 79.7 \\
 & Prompt-BioEL & 91.9 & 94.3 & 82.7 & 89.7 & 72.6 & 89.6 & 86.2 \\
\midrule
\multirow{4}{*}{Generative}
 & BART     & 90.2 & 92.5 & 80.7 & 88.8 & 71.5 & 87.8 & 84.7 \\
 & BioBART  & 89.9 & 93.3 & 81.8 & 89.4 & 71.8 & 88.3 & 85.2 \\
 & GenBioEL & 91.9 & 93.3 & 81.4 & 89.3 & --   & 88.9 & --   \\
 & ANGEL    & 92.8 & \underline{94.5} & 82.8 & \underline{90.2} & \underline{73.3} & 90.0 & \underline{86.7} \\
\midrule
\multirow{4}{*}{LLM-based}
 & GPT-4        & 59.4 & 66.3 & 40.3 & --   & --   & 55.3 & --   \\
 & DeepSeek-R1  & 62.6 & 71.4 & 37.6 & --   & --   & 57.2 & --   \\
 & LLM4BioEL    & \textbf{93.8} & 92.4 & \underline{84.8} & --   & --   & \underline{90.3} & --   \\
 & \textbf{PILOT} & \underline{93.4} & \textbf{95.8} & \textbf{88.0} & \textbf{91.7} & \textbf{74.9} & \textbf{92.4} & \textbf{88.8} \\
\midrule
\multicolumn{2}{c|}{\textit{$\Delta$ vs. best baseline}}
   & $-$0.4 & $+$1.3 & $+$3.2 & $+$1.5 & $+$1.6 & $+$2.1 & $+$2.1 \\
\bottomrule
\end{tabular}
    }
    \caption{Performance (R@1) of all methods across five datasets. ``--'': not reported. \textbf{Bold} and \underline{underline} denote the best and second-best results per column. PILOT achieves the best on most datasets, with a solid lead on average.}
    \label{tab:main_results}
\end{table*}

\begin{table*}[t]
    \centering
    \resizebox{\ifdim\width>\textwidth \textwidth \else \width \fi}{!}{
        \begin{tabular}{l|lllll|l}
\toprule
\textbf{Variant} & \textbf{NCBI} & \textbf{BC5CDR} & \textbf{COMETA} & \textbf{AAP} & \textbf{MM-ST21pv} & \textbf{Avg.} \\
\midrule
\textbf{PILOT} & \textbf{93.4} & \underline{95.8} & \textbf{88.0} & \textbf{91.7} & \underline{74.9} & \textbf{88.8} \\
(1) w/o query reformulation (standardization) & \valu{92.9}{-0.5} & \val{95.7}{-0.1} & \val{86.7}{-1.3} & \valb{91.7}{0.0} & \val{73.4}{-1.5} & \val{88.1}{-0.7} \\
(2) w/o query reformulation (hypernym) & \val{92.6}{-0.8} & \valu{95.8}{0.0} & \valu{87.9}{-0.1} & \valb{91.7}{0.0} & \val{73.9}{-1.0} & \val{88.4}{-0.4} \\
(3) w/o embedding pooling & \val{92.6}{-0.8} & \valb{95.9}{+0.1} & \valb{88.0}{0.0} & \valu{91.5}{-0.2} & \val{74.6}{-0.3} & \valu{88.5}{-0.3} \\
(4) w/o surface-form reranker & \val{88.1}{-5.3} & \val{95.4}{-0.4} & \val{85.2}{-2.8} & \val{90.9}{-0.8} & \val{71.5}{-3.4} & \val{86.2}{-2.6} \\
(5) w/o contextual reranker & \val{91.9}{-1.5} & \val{93.2}{-2.6} & \val{82.7}{-5.3} & \val{81.9}{-9.8} & \val{58.0}{-16.9} & \val{81.5}{-7.3} \\
(6) w/o surface-form negatives & \valb{93.4}{0.0} & \val{95.4}{-0.4} & \val{85.9}{-2.1} & \val{89.6}{-2.1} & \valb{75.2}{+0.3} & \val{87.9}{-0.9} \\
\bottomrule
\end{tabular}
    }
    \caption{Ablation study (R@1) of PILOT. \textbf{Bold} and \underline{underline} denote the best and second-best results per column. Subscripts indicate the difference relative to the full model. PILOT performs best overall, confirming the contribution of each component.}
    \label{tab:ablation_study}
\end{table*}

\subsection{Main results}

Table~\ref{tab:main_results} shows the experimental results across the five datasets. 
PILOT achieves the best result on four of the five and the highest average overall, improving on the strongest prior method ANGEL by 2.1 points. The gains are consistent rather than driven by any single dataset, and are largest on COMETA---a social-media dataset whose irregular surface forms and reliance on context are exactly what the neighborhood-aware, context-sensitive design targets. The only dataset on which PILOT is not first is NCBI, which is a small dataset of 960 test samples and already saturated for all competitive methods. Even so, PILOT ranks a close second by a margin of 0.4 points. 
Since several LLM-based baselines are evaluated only on the first three datasets (NCBI, BC5CDR, and COMETA) and skip the two larger ones (AAP and MM-ST21pv), we additionally report the average over the first three (Avg.\ (first 3)) for a fair comparison, on which PILOT again ranks first, ahead of the second-best LLM4BioEL by 2.1 points.
Overall, PILOT delivers the strongest and most consistent performance across both scientific-literature and social-media settings, and across KBs of widely varying scale.

We further situate PILOT against two of its building blocks that also appear as baselines. First, SapBERT serves as the embedding model for both the retriever and the surface-form reranker. The gap between SapBERT and PILOT is largest on COMETA and MM-ST21pv, the two hardest datasets, showing that PILOT's rerankers are what resolve the ambiguous mentions that pure embedding similarity cannot. Second, PILOT deliberately confines general-purpose LLMs to constrained subtasks---query reformulation and contextual reranking---rather than end-to-end linking. The baseline results for GPT-4 and DeepSeek-R1 justify this choice, as directly prompting them for the answer trails the specialized methods by a wide margin. This confirms that strong general knowledge does not by itself solve entity linking against large biomedical KBs, where the challenge is discriminating among thousands of fine-grained, domain-specific entities that cannot be reliably memorized.

\subsection{Ablation study}
\label{sec:ablation}

We ablate various components across the three stages of PILOT and compare each ablated variant with the full model to validate their contributions (Table~\ref{tab:ablation_study}). Neighborhood-awareness in the retrieval stage comprises three mechanisms---query reformulation for the standardized name, query reformulation for the hypernym, and ontology-based embedding pooling---so rather than removing it wholesale we ablate one at a time by setting $w_s{=}0$, $w_h{=}0$, or $\lambda{=}0$, respectively (rows 1--3). The next two variants are the fusion endpoints of the trained full model: removing the surface-form reranker corresponds to $\alpha{=}1$, and removing the contextual reranker corresponds to $\alpha{=}0$ (rows 4--5). Finally, we remove the $n_\text{sh}$ surface-form negatives from the contextual reranker's training pool and fill the quota entirely from its own hard negatives, by setting $n_\text{sh}{=}0$ while keeping $n$ unchanged (row 6).

We first observe that the three mechanisms in the retrieval stage are indispensable: removing any of them lowers the overall performance by 0.3--0.7 points and degrades results on most datasets. Moreover, each mechanism has its own areas of strength. Standardized-name reformulation helps most on COMETA and MM-ST21pv, hypernym reformulation on NCBI and MM-ST21pv, and embedding pooling on NCBI. The three are therefore complementary, and combining them is what makes the retrieval stage neighborhood-aware across datasets with different characteristics.

Secondly, we notice that using a single reranker without fusion leads to substantial drops. The contextual reranker is the more important one overall, with a drop of 7.3 points against 2.6 points for the surface-form reranker. Per dataset, however, the picture varies: the surface-form reranker matters more on NCBI, while the contextual reranker matters more on the others. The two are thus complementary---one sharpening surface-form discrimination and the other performing contextual disambiguation---and only their fusion yields accurate and robust performance across all datasets. Additionally, on MM-ST21pv, the hardest dataset, the contextual reranker alone contributes 16.9 points, suggesting that annotation conventions may be a major reason for the low performance there and that training on in-domain data to capture these conventions is the most effective lever on such datasets.

We finally find that surface-form negatives are necessary for training the contextual reranker. Removing them and relying solely on the reranker's own hard negatives costs 0.9 points overall, with the largest drops of 2.1 points on both COMETA and AAP. This shows that although the contextual reranker disambiguates mentions from context, lexically similar candidates can still confuse it and should be explicitly contrasted during training.

\begin{figure*}[t]
    \centering
    \begin{subfigure}{\textwidth}
        \centering
        \includegraphics[width=\linewidth]{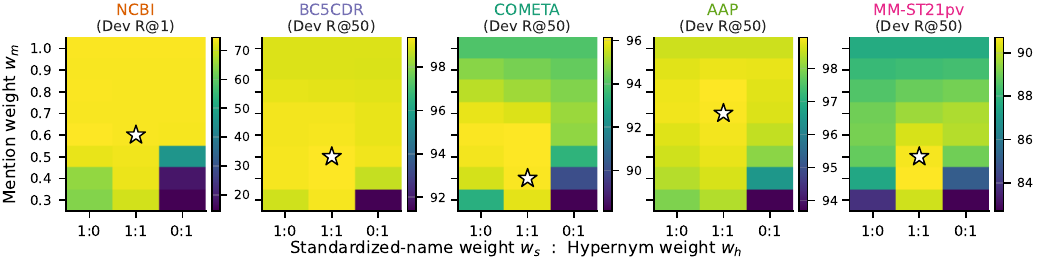}
        \caption{Neighborhood-aware retriever. We report R@1 for NCBI, as its R@50 is saturated near 100\%.}
        \label{fig:sensitivity_retrieval_mix}
    \end{subfigure}
    \begin{subfigure}{\textwidth}
        \centering
        \includegraphics[width=\linewidth]{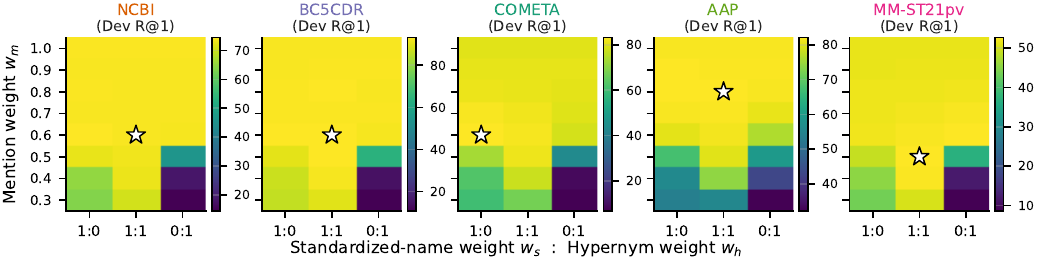}
        \caption{Surface-form reranker.}
        \label{fig:sensitivity_surfaceform_mix}
    \end{subfigure}
    \caption{PILOT's sensitivity to the query-mix weights $(w_m, w_s, w_h)$ on the development set. $\bigstar$ marks the optimal (selected) setting per dataset. PILOT is robust to hyperparameters near the optimum.
    }
    \label{fig:sensitivity_mix}
\end{figure*}

\subsection{Hyperparameter sensitivity analysis}

We conduct extensive sensitivity analyses on the development set to examine how PILOT responds to its key hyperparameters. We first study the tunable ones: the query-mix weights $(w_m, w_s, w_h)$ (Figure~\ref{fig:sensitivity_mix}), the pooling weight $\lambda$ (Figure~\ref{fig:sensitivity_pool}), and the fusion weight $\alpha$ (Figure~\ref{fig:sensitivity_alpha}). We then turn to the hyperparameters that we keep fixed because they affect training and are expensive to tune: the retrieval depth $K$, the number of hard negatives $n$, and the number of training epochs $E$ (Figure~\ref{fig:sensitivity_fixed}). Overall, PILOT is robust to all of them, with performance changing little near the selected optimum.

For query reformulation, the standardized-name and hypernym reformulations are in most cases best used together (Figure~\ref{fig:sensitivity_mix}), echoing the ablation study (rows 1--2, Table~\ref{tab:ablation_study}). How much weight they receive relative to the mention itself, however, is dataset-dependent, with the optimum falling in a rough range of $w_m{\in}[0.4,0.8]$.

Pooling behaves very differently at the two stages it is applied to (Figure~\ref{fig:sensitivity_pool}). It barely moves the retriever, whose development R@50 stays nearly flat across $\lambda$, but has a pronounced effect on the surface-form reranker, whose development R@1 varies considerably. This is expected: the retriever only needs the gold entity to fall within the top $K$, whereas the reranker must place it first. On COMETA the contrast is sharpest, as the best pooling weight for the retriever is the worst one for the surface-form reranker. Hence, though the surface-form reranker is a second pass of the retriever, their hyperparameters should be tuned separately to serve their respective goals of recall and precision.

Fusion follows an inverted-U pattern (Figure~\ref{fig:sensitivity_alpha}): performance rises with $\alpha$ and then falls, and is worst at the two endpoints, where either the contextual or the surface-form reranker is removed, again corroborating the ablation study (rows 4--5, Table~\ref{tab:ablation_study}). The optima cluster in $[0.35,0.65]$, indicating that both rerankers contribute substantially to the final ranking and confirming that fusion is necessary rather than cosmetic.

Turning to the fixed hyperparameters (Figure~\ref{fig:sensitivity_fixed}), we observe three trends. First, performance improves with the retrieval depth and saturates at $K{=}50$, beyond which deeper retrieval adds linear inference cost for no further gain, and the benefit of a large $K$ is most pronounced on the harder datasets---COMETA ($+1.6$), AAP ($+1.1$), and MM-ST21pv ($+1.0$). Such a depth is out of reach for competitive prior methods such as Prompt-BioEL ($K{=}6$) and LLM4BioEL ($K{\in}[1,10]$), and is exactly what our pointwise reranker is designed to support. Second, raising the number of hard negatives from 4 to 8 gives a clear improvement, while going further buys little at a linearly growing training cost. Third, performance jumps sharply over the first training epoch, underlining the need to train the reranker to capture the annotation conventions, but improves much more slowly over the second, showing that two epochs suffice. These trends confirm that our fixed choices of $K{=}50$, $n{=}8$, and $E{=}2$ are well justified.

\subsection{Throughput}

\begin{table}[t]
    \centering
    \resizebox{\ifdim\width>\columnwidth \columnwidth \else \width \fi}{!}{
        \begin{tabular}{l|ccccc}
\toprule
\textbf{Method} & \textbf{NCBI} & \textbf{BC5CDR} & \textbf{COMETA} & \textbf{AAP} & \textbf{MM-ST21pv} \\
\midrule
LLM4BioEL & 0.20 & 0.26 & 0.26 & --  & --  \\
\textbf{PILOT}      & 1.90 & 1.81 & 2.81 & 4.27 & 5.23 \\
\midrule
\textit{Ratio} & \text9.50$\times$ & 6.96$\times$ & 10.81$\times$ & -- & -- \\
\bottomrule
\end{tabular}
    }
    \caption{Throughput (mentions per GPU-second) of PILOT and LLM4BioEL, the strongest LLM-based baseline. \textit{Ratio} = PILOT / LLM4BioEL.
    }
    \label{tab:throughput}
\end{table}

To examine whether a large retrieval depth would hurt throughput, we compare PILOT with the strongest LLM-based baseline, LLM4BioEL (Table~\ref{tab:throughput}). With the 4B model on NCBI, BC5CDR, COMETA, and AAP, PILOT reaches 1.81--4.27 mentions per GPU-second, mostly depending on the length of the context. On MM-ST21pv, PILOT uses the smaller 0.6B model and attains an even higher throughput of 5.23, while still delivering the best R@1 among all methods (Table~\ref{tab:main_results}). PILOT is therefore deployable for time-sensitive real-world inference despite its large retrieval depth, thanks to its parallelizable pointwise reranker. Compared with LLM4BioEL, PILOT is 7--11$\times$ faster. This speedup comes from a much smaller LLM (0.6B/4B vs.\ 70B) and from not relying on in-context learning, whereas LLM4BioEL requires up to 80 few-shot examples in its prompts.

\section{Conclusion}

We presented PILOT, a retrieve-and-rerank framework for biomedical entity linking that considers ontological neighborhood and mention context. Across five benchmarks PILOT achieves the state of the art on average while remaining efficient at inference. Our analyses point to three findings. Injecting ontological structure helps retrieval most when it is applied to both the query and the KB. The surface-form and contextual views are complementary for reranking. Scoring candidates independently is what makes a deep candidate pool affordable, and depth is where much of the remaining accuracy lies on the hardest datasets.

\section*{Limitations}

\paragraph{Dependence on ontological structure.} Both directions of our neighborhood-aware retrieval assume a usable ``is-a'' hierarchy in the target KB, which may be incomplete or unavailable for some knowledge bases. In such cases the neighborhood-aware mechanisms would provide little benefit.

\paragraph{Supervision requirements.} PILOT trains the retriever and the contextual reranker on in-domain data and selects the query-mix, pooling, and fusion weights on a labeled development set for each dataset. This is what allows the model to absorb dataset-specific annotation conventions, but it also means the reported numbers do not speak to the zero-shot or cross-corpus setting, where no in-domain supervision is available.

\paragraph{Scope of evaluation.} All five benchmarks are English and generally biomedical. Whether PILOT transfers to other languages or domains remains untested. Also, all benchmarks come from scientific literature or social media, so PILOT's performance in real-world clinical settings is unknown.


\bibliography{custom}

@inproceedings{sung2020biomedical,
  title={Biomedical entity representations with synonym marginalization},
  author={Sung, Mujeen and Jeon, Hwisang and Lee, Jinhyuk and Kang, Jaewoo},
  booktitle={Proceedings of the 58th annual meeting of the association for computational linguistics},
  pages={3641--3650},
  year={2020}
}

@inproceedings{liu2021self,
  title={Self-alignment pretraining for biomedical entity representations},
  author={Liu, Fangyu and Shareghi, Ehsan and Meng, Zaiqiao and Basaldella, Marco and Collier, Nigel},
  booktitle={Proceedings of the 2021 conference of the North American chapter of the association for computational linguistics: human language technologies},
  pages={4228--4238},
  year={2021}
}

@inproceedings{lai2021bert,
  title={BERT might be overkill: A tiny but effective biomedical entity linker based on residual convolutional neural networks},
  author={Lai, Tuan and Ji, Heng and Zhai, ChengXiang},
  booktitle={Findings of the Association for Computational Linguistics: EMNLP 2021},
  pages={1631--1639},
  year={2021}
}

@inproceedings{zhang2022knowledge,
  title={Knowledge-rich self-supervision for biomedical entity linking},
  author={Zhang, Sheng and Cheng, Hao and Vashishth, Shikhar and Wong, Cliff and Xiao, Jinfeng and Liu, Xiaodong and Naumann, Tristan and Gao, Jianfeng and Poon, Hoifung},
  booktitle={Findings of the Association for Computational Linguistics: EMNLP 2022},
  pages={868--880},
  year={2022}
}

@inproceedings{xu2023improving,
  title={Improving biomedical entity linking with cross-entity interaction},
  author={Xu, Zhenran and Chen, Yulin and Hu, Baotian},
  booktitle={Proceedings of the AAAI Conference on Artificial Intelligence},
  volume={37},
  number={11},
  pages={13869--13877},
  year={2023}
}

@inproceedings{lewis2020bart,
  title={BART: Denoising sequence-to-sequence pre-training for natural language generation, translation, and comprehension},
  author={Lewis, Mike and Liu, Yinhan and Goyal, Naman and Ghazvininejad, Marjan and Mohamed, Abdelrahman and Levy, Omer and Stoyanov, Veselin and Zettlemoyer, Luke},
  booktitle={Proceedings of the 58th annual meeting of the association for computational linguistics},
  pages={7871--7880},
  year={2020}
}

@inproceedings{yuan2022biobart,
  title={BioBART: Pretraining and evaluation of a biomedical generative language model},
  author={Yuan, Hongyi and Yuan, Zheng and Gan, Ruyi and Zhang, Jiaxing and Xie, Yutao and Yu, Sheng},
  booktitle={Proceedings of the 21st Workshop on Biomedical Language Processing},
  pages={97--109},
  year={2022}
}

@inproceedings{yuan2022generative,
  title={Generative biomedical entity linking via knowledge base-guided pre-training and synonyms-aware fine-tuning},
  author={Yuan, Hongyi and Yuan, Zheng and Yu, Sheng},
  booktitle={Proceedings of the 2022 Conference of the North American Chapter of the Association for Computational Linguistics: Human Language Technologies},
  pages={4038--4048},
  year={2022}
}

@inproceedings{kim2025learning,
  title={Learning from negative samples in biomedical generative entity linking},
  author={Kim, Chanhwi and Kim, Hyunjae and Park, Sihyeon and Lee, Jiwoo and Sung, Mujeen and Kang, Jaewoo},
  booktitle={Findings of the Association for Computational Linguistics: ACL 2025},
  pages={10714--10730},
  year={2025}
}

@article{achiam2023gpt,
  title={Gpt-4 technical report},
  author={Achiam, Josh and Adler, Steven and Agarwal, Sandhini and Ahmad, Lama and Akkaya, Ilge and Aleman, Florencia Leoni and Almeida, Diogo and Altenschmidt, Janko and Altman, Sam and Anadkat, Shyamal and others},
  journal={arXiv preprint arXiv:2303.08774},
  year={2023}
}

@article{guo2025deepseek,
  title={Deepseek-r1: Incentivizing reasoning capability in llms via reinforcement learning},
  author={Guo, Daya and Yang, Dejian and Zhang, Haowei and Song, Junxiao and Wang, Peiyi and Zhu, Qihao and Xu, Runxin and Zhang, Ruoyu and Ma, Shirong and Bi, Xiao and others},
  journal={arXiv preprint arXiv:2501.12948},
  year={2025}
}

@article{lin2025guiding,
  title={Guiding Large Language Models for Biomedical Entity Linking via Restrictive and Contrastive Decoding},
  author={Lin, Zhenxi and Zhang, Ziheng and Wu, Jian and Zheng, Yefeng and Wu, Xian},
  journal={Proceedings of the Findings of the Association for Computational Linguistics: EMNLP},
  year={2025}
}

@article{dougan2014ncbi,
  title={NCBI disease corpus: a resource for disease name recognition and concept normalization},
  author={Do{\u{g}}an, Rezarta Islamaj and Leaman, Robert and Lu, Zhiyong},
  journal={Journal of biomedical informatics},
  volume={47},
  pages={1--10},
  year={2014},
  publisher={Elsevier}
}

@article{li2016biocreative,
  title={BioCreative V CDR task corpus: a resource for chemical disease relation extraction},
  author={Li, Jiao and Sun, Yueping and Johnson, Robin J and Sciaky, Daniela and Wei, Chih-Hsuan and Leaman, Robert and Davis, Allan Peter and Mattingly, Carolyn J and Wiegers, Thomas C and Lu, Zhiyong},
  journal={Database},
  volume={2016},
  year={2016},
  publisher={Oxford Academic}
}

@inproceedings{basaldella2020cometa,
  title={COMETA: A corpus for medical entity linking in the social media},
  author={Basaldella, Marco and Liu, Fangyu and Shareghi, Ehsan and Collier, Nigel},
  booktitle={Proceedings of the 2020 Conference on Empirical Methods in Natural Language Processing (EMNLP)},
  pages={3122--3137},
  year={2020}
}

@inproceedings{limsopatham2016normalising,
  title={Normalising medical concepts in social media texts by learning semantic representation},
  author={Limsopatham, Nut and Collier, Nigel},
  booktitle={Proceedings of the 54th Annual Meeting of the Association for Computational Linguistics (volume 1: long papers)},
  pages={1014--1023},
  year={2016}
}

@article{mohan2019medmentions,
  title={Medmentions: A large biomedical corpus annotated with umls concepts},
  author={Mohan, Sunil and Li, Donghui},
  journal={arXiv preprint arXiv:1902.09476},
  year={2019}
}

@article{lipscomb2000medical,
  title={Medical subject headings (MeSH)},
  author={Lipscomb, Carolyn E},
  journal={Bulletin of the Medical Library Association},
  volume={88},
  number={3},
  pages={265},
  year={2000}
}

@article{hamosh2005online,
  title={Online Mendelian Inheritance in Man (OMIM), a knowledgebase of human genes and genetic disorders},
  author={Hamosh, Ada and Scott, Alan F and Amberger, Joanna S and Bocchini, Carol A and McKusick, Victor A},
  journal={Nucleic acids research},
  volume={33},
  number={suppl\_1},
  pages={D514--D517},
  year={2005},
  publisher={Oxford University Press}
}

@article{davis2012medic,
  title={MEDIC: a practical disease vocabulary used at the Comparative Toxicogenomics Database},
  author={Davis, Allan Peter and Wiegers, Thomas C and Rosenstein, Michael C and Mattingly, Carolyn J},
  journal={Database},
  volume={2012},
  pages={bar065},
  year={2012},
  publisher={Oxford University Press}
}

@article{chang2021use,
  title={The use of SNOMED CT, 2013-2020: a literature review},
  author={Chang, Eunsuk and Mostafa, Javed},
  journal={Journal of the American Medical Informatics Association},
  volume={28},
  number={9},
  pages={2017--2026},
  year={2021},
  publisher={Oxford University Press}
}

@article{bodenreider2004unified,
  title={The unified medical language system (UMLS): integrating biomedical terminology},
  author={Bodenreider, Olivier},
  journal={Nucleic acids research},
  volume={32},
  number={suppl\_1},
  pages={D267--D270},
  year={2004},
  publisher={Oxford University Press}
}

@article{rocchio1971relevance,
  title={Relevance feedback in information retrieval},
  author={Rocchio Jr, Joseph John},
  journal={The SMART retrieval system: experiments in automatic document processing},
  year={1971},
  publisher={Englewood Cliffs}
}

@inproceedings{qin2024large,
  title={Large language models are effective text rankers with pairwise ranking prompting},
  author={Qin, Zhen and Jagerman, Rolf and Hui, Kai and Zhuang, Honglei and Wu, Junru and Yan, Le and Shen, Jiaming and Liu, Tianqi and Liu, Jialu and Metzler, Donald and others},
  booktitle={Findings of the Association for Computational Linguistics: NAACL 2024},
  pages={1504--1518},
  year={2024}
}

@article{liu2024lost,
  title={Lost in the middle: How language models use long contexts},
  author={Liu, Nelson F and Lin, Kevin and Hewitt, John and Paranjape, Ashwin and Bevilacqua, Michele and Petroni, Fabio and Liang, Percy},
  journal={Transactions of the association for computational linguistics},
  volume={12},
  pages={157--173},
  year={2024}
}

@inproceedings{sun2023chatgpt,
  title={Is ChatGPT good at search? investigating large language models as re-ranking agents},
  author={Sun, Weiwei and Yan, Lingyong and Ma, Xinyu and Wang, Shuaiqiang and Ren, Pengjie and Chen, Zhumin and Yin, Dawei and Ren, Zhaochun},
  booktitle={Proceedings of the 2023 conference on empirical methods in natural language processing},
  pages={14918--14937},
  year={2023}
}

@article{yang2025qwen3,
  title={Qwen3 technical report},
  author={Yang, An and Li, Anfeng and Yang, Baosong and Zhang, Beichen and Hui, Binyuan and Zheng, Bo and Yu, Bowen and Gao, Chang and Huang, Chengen and Lv, Chenxu and others},
  journal={arXiv preprint arXiv:2505.09388},
  year={2025}
}

@article{zhang2025qwen3,
  title={Qwen3 embedding: Advancing text embedding and reranking through foundation models},
  author={Zhang, Yanzhao and Li, Mingxin and Long, Dingkun and Zhang, Xin and Lin, Huan and Yang, Baosong and Xie, Pengjun and Yang, An and Liu, Dayiheng and Lin, Junyang and others},
  journal={arXiv preprint arXiv:2506.05176},
  year={2025}
}

@article{hu2022lora,
  title={Lora: Low-rank adaptation of large language models.},
  author={Hu, Edward J and Shen, Yelong and Wallis, Phillip and Allen-Zhu, Zeyuan and Li, Yuanzhi and Wang, Shean and Wang, Liang and Chen, Weizhu and others},
  journal={Iclr},
  volume={1},
  number={2},
  pages={3},
  year={2022}
}

@inproceedings{zhuang2024beyond,
  title={Beyond yes and no: Improving zero-shot llm rankers via scoring fine-grained relevance labels},
  author={Zhuang, Honglei and Qin, Zhen and Hui, Kai and Wu, Junru and Yan, Le and Wang, Xuanhui and Bendersky, Michael},
  booktitle={Proceedings of the 2024 conference of the North American chapter of the Association for Computational Linguistics: Human language technologies (volume 2: short papers)},
  pages={358--370},
  year={2024}
}

@article{nogueira2019multi,
  title={Multi-stage document ranking with BERT},
  author={Nogueira, Rodrigo and Yang, Wei and Cho, Kyunghyun and Lin, Jimmy},
  journal={arXiv preprint arXiv:1910.14424},
  year={2019}
}

@inproceedings{nogueira2020document,
  title={Document ranking with a pretrained sequence-to-sequence model},
  author={Nogueira, Rodrigo and Jiang, Zhiying and Pradeep, Ronak and Lin, Jimmy},
  booktitle={Findings of the association for computational linguistics: EMNLP 2020},
  pages={708--718},
  year={2020}
}

@inproceedings{zhuang2023rankt5,
  title={Rankt5: Fine-tuning t5 for text ranking with ranking losses},
  author={Zhuang, Honglei and Qin, Zhen and Jagerman, Rolf and Hui, Kai and Ma, Ji and Lu, Jing and Ni, Jianmo and Wang, Xuanhui and Bendersky, Michael},
  booktitle={Proceedings of the 46th international ACM SIGIR conference on research and development in information retrieval},
  pages={2308--2313},
  year={2023}
}

@inproceedings{zhuang2024setwise,
  title={A setwise approach for effective and highly efficient zero-shot ranking with large language models},
  author={Zhuang, Shengyao and Zhuang, Honglei and Koopman, Bevan and Zuccon, Guido},
  booktitle={Proceedings of the 47th International ACM SIGIR Conference on Research and Development in Information Retrieval},
  pages={38--47},
  year={2024}
}

@inproceedings{luo2024prp,
  title={Prp-graph: Pairwise ranking prompting to llms with graph aggregation for effective text re-ranking},
  author={Luo, Jian and Chen, Xuanang and He, Ben and Sun, Le},
  booktitle={Proceedings of the 62nd annual meeting of the association for computational linguistics (volume 1: long papers)},
  pages={5766--5776},
  year={2024}
}

@article{pradeep2023rankzephyr,
  title={Rankzephyr: Effective and robust zero-shot listwise reranking is a breeze!},
  author={Pradeep, Ronak and Sharifymoghaddam, Sahel and Lin, Jimmy},
  journal={arXiv preprint arXiv:2312.02724},
  year={2023}
}

@inproceedings{wu2022conqrr,
  title={Conqrr: Conversational query rewriting for retrieval with reinforcement learning},
  author={Wu, Zeqiu and Luan, Yi and Rashkin, Hannah and Reitter, David and Hajishirzi, Hannaneh and Ostendorf, Mari and Tomar, Gaurav Singh},
  booktitle={Proceedings of the 2022 Conference on Empirical Methods in Natural Language Processing},
  pages={10000--10014},
  year={2022}
}

@inproceedings{mao2023large,
  title={Large language models know your contextual search intent: A prompting framework for conversational search},
  author={Mao, Kelong and Dou, Zhicheng and Mo, Fengran and Hou, Jiewen and Chen, Haonan and Qian, Hongjin},
  booktitle={Findings of the Association for Computational Linguistics: EMNLP 2023},
  pages={1211--1225},
  year={2023}
}

@inproceedings{ma2023query,
  title={Query rewriting in retrieval-augmented large language models},
  author={Ma, Xinbei and Gong, Yeyun and He, Pengcheng and Zhao, Hai and Duan, Nan},
  booktitle={Proceedings of the 2023 conference on empirical methods in natural language processing},
  pages={5303--5315},
  year={2023}
}

@inproceedings{mao2024rafe,
  title={RaFe: Ranking feedback improves query rewriting for RAG},
  author={Mao, Shengyu and Jiang, Yong and Chen, Boli and Li, Xiao and Wang, Peng and Wang, Xinyu and Xie, Pengjun and Huang, Fei and Chen, Huajun and Zhang, Ningyu},
  booktitle={Findings of the Association for Computational Linguistics: EMNLP 2024},
  pages={884--901},
  year={2024}
}

@inproceedings{peng2024large,
  title={Large language model based long-tail query rewriting in taobao search},
  author={Peng, Wenjun and Li, Guiyang and Jiang, Yue and Wang, Zilong and Ou, Dan and Zeng, Xiaoyi and Xu, Derong and Xu, Tong and Chen, Enhong},
  booktitle={Companion Proceedings of the ACM Web Conference 2024},
  pages={20--28},
  year={2024}
}

@inproceedings{wang2023query2doc,
  title={Query2doc: Query expansion with large language models},
  author={Wang, Liang and Yang, Nan and Wei, Furu},
  booktitle={Proceedings of the 2023 Conference on Empirical Methods in Natural Language Processing},
  pages={9414--9423},
  year={2023}
}

@inproceedings{gao2023precise,
  title={Precise zero-shot dense retrieval without relevance labels},
  author={Gao, Luyu and Ma, Xueguang and Lin, Jimmy and Callan, Jamie},
  booktitle={Proceedings of the 61st Annual Meeting of the Association for Computational Linguistics (Volume 1: Long Papers)},
  pages={1762--1777},
  year={2023}
}

@inproceedings{lei2024corpus,
  title={Corpus-steered query expansion with large language models},
  author={Lei, Yibin and Cao, Yu and Zhou, Tianyi and Shen, Tao and Yates, Andrew},
  booktitle={Proceedings of the 18th Conference of the European Chapter of the Association for Computational Linguistics (Volume 2: Short Papers)},
  pages={393--401},
  year={2024}
}

@inproceedings{decao2021autoregressive,
  author    = {Nicola {De Cao} and
               Gautier Izacard and
               Sebastian Riedel and
               Fabio Petroni},
  title     = {Autoregressive Entity Retrieval},
  booktitle = {9th International Conference on Learning Representations, {ICLR} 2021,
               Virtual Event, Austria, May 3-7, 2021},
  publisher = {OpenReview.net},
  year      = {2021},
  url       = {https://openreview.net/forum?id=5k8F6UU39V},
}

@article{huang2024building,
  title={Building a literature knowledge base towards transparent biomedical AI},
  author={Huang, Yuanhao and Han, Zhaowei and Luo, Xin and Luo, Xuteng and Gao, Yijia and Zhao, Meiqi and Tang, Feitong and Wang, Yiqun and Chen, Jiyu and Li, Chengfan and others},
  journal={bioRxiv},
  pages={2024--09},
  year={2024},
  publisher={Cold Spring Harbor Laboratory}
}

@article{tao2026autopcr,
  title={AutoPCR: automated phenotype concept recognition by prompting},
  author={Tao, Yicheng and Huang, Yuanhao and Wang, Yiqun and Luo, Xin and Liu, Jie},
  journal={Bioinformatics},
  volume={42},
  number={Supplement\_1},
  pages={btag304},
  year={2026},
  publisher={Oxford University Press}
}

\appendix

\begin{table*}[t]
    \centering
    \resizebox{\ifdim\width>\textwidth \textwidth \else \width \fi}{!}{
        \begin{tabular}{lllrrr}
\toprule
\textbf{Dataset} & \textbf{Entity types} & \textbf{Train / Dev / Test} & \textbf{Target KB} & \textbf{\# Entities} \\
\midrule
NCBI      & Disease           & 5,784 / 787 / 960        & MEDIC          & 14,944    \\
BC5CDR    & Disease \& chemical   & 9,285 / 9,515 / 9,654    & MeSH           & 268,162   \\
COMETA    & Biomedical   & 13,489 / 2,176 / 4,350   & SNOMED CT      & 350,830   \\
AAP       & Medical   & 15,665 / 793 / 866       & SNOMED CT         & 1,036     \\
MM-ST21pv & 21 UMLS types      & 121,498 / 40,600 / 39,922 & UMLS          & 3,092,324 \\
\bottomrule
\end{tabular}
    }
    \caption{Statistics of the used datasets and their target KBs. For AAP, only the subset of SNOMED CT that contains AAP mentions is considered, following the ANGEL work.}
    \label{tab:dataset_stats}
\end{table*}

\begin{figure*}[t]
    \centering
    \begin{subfigure}{0.45\textwidth}
        \centering
        \includegraphics[width=\linewidth]{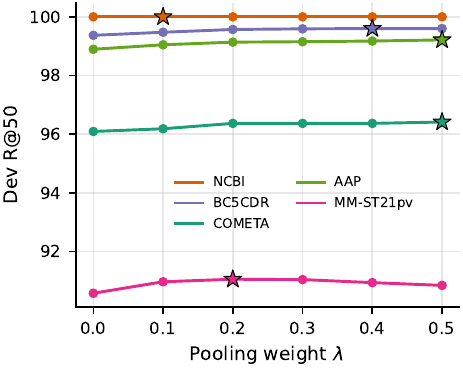}
        \caption{Neighborhood-aware retriever.}
        \label{fig:sensitivity_retrieval_pool}
    \end{subfigure}
    \hfill
    \begin{subfigure}{0.45\textwidth}
        \centering
        \includegraphics[width=\linewidth]{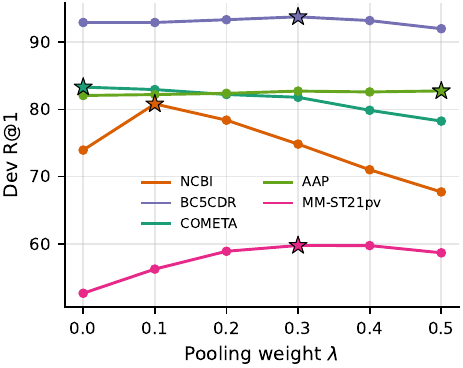}
        \caption{Surface-form reranker.}
        \label{fig:sensitivity_surfaceform_pool}
    \end{subfigure}
    \caption{PILOT's sensitivity to the pooling weight $\lambda$ on the development set. $\bigstar$ marks the optimal (selected) setting per dataset. PILOT is robust to hyperparameters near the optimum.}
    \label{fig:sensitivity_pool}
\end{figure*}

\begin{figure}[t]
  \centering
  \includegraphics[width=\columnwidth]{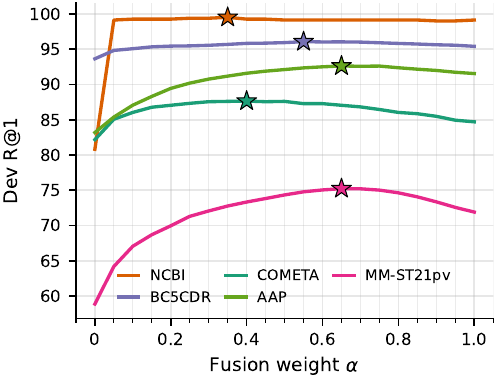}
  \caption{PILOT's sensitivity to the fusion weight $\alpha$ on the development set. $\bigstar$ marks the optimal (selected) setting per dataset. PILOT is robust to hyperparameters near the optimum.
  }
  \label{fig:sensitivity_alpha}
\end{figure}

\begin{figure*}[t]
  \centering
  \includegraphics[width=\textwidth]{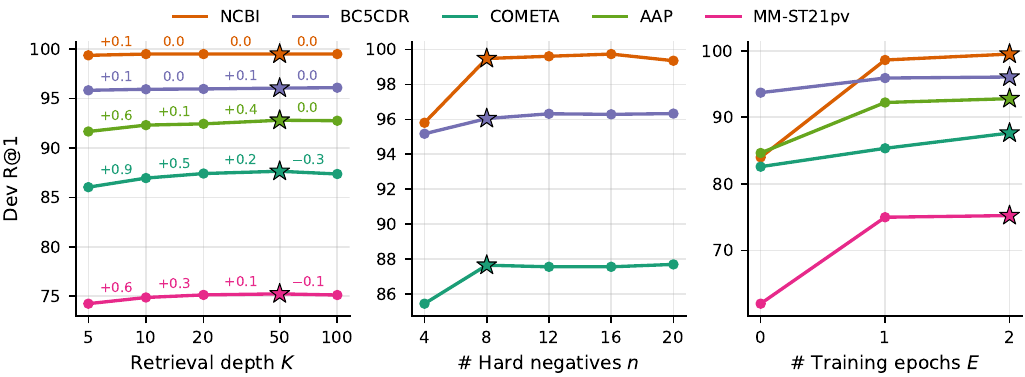}
  \caption{PILOT's sensitivity to the retrieval depth $K$, the number of hard negatives $n$, and the number of training epochs $E$ on the development set. AAP and MM-ST21pv are excluded from the sweep over $n$ due to the heavy computational cost. $\bigstar$ marks the fixed setting per dataset, at which PILOT performs near-optimally.}
  \label{fig:sensitivity_fixed}
\end{figure*}

\section{Prompts}

\subsection{Generative query reformulation prompts}
\label{app:query_prompts}

Query reformulation issues two independent generations per mention using the following prompts.

\paragraph{Standardized name.}
\begin{quote}
    \textbf{[System]}\\
    You are a biomedical normalization assistant. Given a mention from a clinical/biomedical text, return the most appropriate standard scientific/medical name for the concept the mention refers to (e.g., the canonical UMLS or SNOMED CT preferred term). Reply with ONLY the standard name, no explanation, no punctuation, no list.
\end{quote}
\begin{quote}
    \textbf{[User]}\\
    Sentence: ``\{context\}''\\
    Mention: ``\{mention\}''\\
    Standard scientific name:
\end{quote}

\paragraph{Hypernym.}
\begin{quote}
    \textbf{[System]}\\
    You are a biomedical ontology assistant. Given a mention from clinical/biomedical text, first expand any abbreviation, then name the BROADER parent category or general concept it is a kind of (its hypernym in UMLS/SNOMED CT) — e.g. ``apoptotic cells'' → ``cell'', ``CABG'' → ``coronary artery bypass''. Reply with ONLY a short general term, no explanation, no punctuation.
\end{quote}
\begin{quote}
    \textbf{[User]}\\
    Sentence: ``\{context\}''\\
    Mention: ``\{mention\}''\\
    Broader category (hypernym):
\end{quote}

\subsection{Contextual reranker prompt}
\label{app:reranker_prompt}

The contextual reranker uses the default system prompt and a customized user prompt:
\begin{quote}
    <Instruct>: Given a biomedical entity mention in its sentence context, determine whether the candidate concept from the knowledge base is its correct standardized form.\\
    <Query>: In the sentence ``\{context\}'', the mention ``\{mention\}'' refers to:\\
    <Document>: \{candidate document\}
\end{quote}
where the candidate document is the candidate's synonyms followed by its one-hop ontology suffix (up to three neighbor names per direction), e.g.:

\begin{quote}
    hypertensive disorder (synonyms: high blood pressure, HTN) [parents: cardiovascular disease] [children: essential hypertension; secondary hypertension; malignant hypertension] [siblings: hypotension]
\end{quote}

\section{Implementation Details on AAP}
\label{sec:details_aap}

\begin{table}[t]
    \centering
    \resizebox{\ifdim\width>\columnwidth \columnwidth \else \width \fi}{!}{
        \begin{tabular}{c| cccc cccc c}
\toprule
\multirow{2}{*}{\textbf{Fold}} & \multicolumn{4}{c}{\textbf{Retriever}} & \multicolumn{4}{c}{\textbf{Reranker}} & \multirow{2}{*}{$\alpha$} \\
\cmidrule(lr){2-5} \cmidrule(lr){6-9}
 & $w_m$ & $w_s$ & $w_h$ & $\lambda$ & $w_m$ & $w_s$ & $w_h$ & $\lambda$ & \\
\midrule
0 & .8 & .20 & .00 & .3 & .9 & .00 & .10 & .2 & .65 \\
1 & .8 & .10 & .10 & .3 & .8 & .10 & .10 & .5 & .70 \\
2 & .6 & .20 & .20 & .1 & .7 & .15 & .15 & .4 & .60 \\
3 & .7 & .15 & .15 & .2 & .9 & .05 & .05 & .5 & .70 \\
4 & .7 & .15 & .15 & .5 & .8 & .20 & .00 & .3 & .65 \\
5 & .6 & .20 & .20 & .3 & .8 & .20 & .00 & .4 & .55 \\
6 & .8 & .20 & .00 & .4 & .7 & .30 & .00 & .5 & .75 \\
7 & .8 & .10 & .10 & .5 & .8 & .00 & .20 & .3 & .90 \\
8 & .7 & .15 & .15 & .5 & .8 & .00 & .20 & .4 & .65 \\
9 & .9 & .10 & .00 & .5 & .8 & .00 & .20 & .5 & .60 \\
\midrule
\textbf{Avg.} & .74 & .16 & .11 & .36 & .80 & .10 & .10 & .40 & .68 \\
\bottomrule
\end{tabular}
    }
    \caption{Per-fold hyperparameter values on AAP.}
    \label{tab:details_aap}
\end{table}

AAP is evaluated via 10-fold cross-validation, where each fold provides independent train, development, and test splits. Accordingly, the query-mix, pooling, and fusion weights are selected separately on each fold's own development set. For clarity of presentation, the hyperparameter values reported in the main text (Figures~\ref{fig:sensitivity_mix},~\ref{fig:sensitivity_pool}, and~\ref{fig:sensitivity_alpha}) are the uniform values shared across folds. The actual per-fold selected values are listed in Table~\ref{tab:details_aap}. On average, these per-fold values remain close to the uniform ones---query-mix weights differ by at most 0.05 and fusion weights by at most 0.15.

\end{document}